\documentclass{ws-rv9x6}
\usepackage{ws-rv-van}             
\usepackage{graphicx}
\usepackage{slashed}

\newcommand{\eq}{\begin{equation}}
\newcommand{\eqe}{\end{equation}}
\newcommand{\be}{\begin{equation}}
\newcommand{\ee}{\end{equation}}
\newcommand{\g}{\gamma}

\newcommand{\eqa}{\begin{eqnarray}}
\newcommand{\eqae}{\end{eqnarray}}
\newcommand{\bea}{\begin{eqnarray}}
\newcommand{\eea}{\end{eqnarray}}
\newcommand{\bma}{\begin{matrix}}
\newcommand{\ema}{\end{matrix}}
\newcommand{\bpm}{\begin{pmatrix}}
\newcommand{\epm}{\end{pmatrix}}
\newcommand{\nn}{\nonumber}

\newcommand{\w}{\omega}
\newcommand{\e}{\epsilon}

\def\0{\over } \def\1{\vec } \def\2{{1\over2}} \def\4{{1\over4}}
\def\5{\bar } 
\def\6{\partial }
\def\7#1{\slashed{#1}}
\def\8#1{{\textstyle{#1}}} \def\9#1{{\underline{#1}}} 
\def\3#1{{\bf #1}}
\def\nn{\nonumber\\ }
\def\intx{\int\!d^3x\, }

\def\olrp{\hbox{\raisebox{2ex}{$\scriptstyle \leftrightarrow$}} \llap{$\partial$}}

\begin{document}
\setcounter{chapter}{1}
\chapter*{Quantum corrections to solitons and BPS saturation$^1$}{\footnotetext[1]{Contribution
to ``Fundamental Interactions---A Memorial Volume for Wolfgang Kummer'',
D.\ Grumiller, A.\ Rebhan, D.V.\ Vassilevich (eds.)}}
\author{A. Rebhan}
\address{Institut f\"ur Theoretische Physik\\ Technische Universit\"at Wien, 
A-1040 Wien, Austria \\ rebhana@tph.tuwien.ac.at}
\author{P. van Nieuwenhuizen}
\address{C.N. Yang Institute for Theoretical Physics\\ 
{Stony Brook University, Stony Brook, NY 11794-3840, USA}\\
{vannieu@insti.physics.sunysb.edu}}  
\author[A. Rebhan, P. van Nieuwenhuizen, R. Wimmer]{R. Wimmer}
\address{Laboratoire de Physique, ENS Lyon,\\
46 all\'ee d'Italie, F-69364 Lyon CEDEX 07, France\\
robert.wimmer@ens-lyon.fr}
\begin{abstract}
We review our work of the past decade on one-loop quantum corrections to the mass $M$ and central charge $Z$ of solitons in supersymmetric field theories: the kink, the vortex, and the 
monopoles (focussing on the kink and the monopoles here). In each case a new feature was needed to obtain BPS saturation: a new anomaly-like contribution to $Z$ for the kink and the $N=2$ monopole, the effect of classical winding of the quantum vortex contributing to $Z$, surface terms contributing to $M$ of the $N=4$ monopole and to $Z$ of the $N=2$ and $N=4$ monopoles, and composite operator renormalization for the currents of the ``finite" $N=4$ model. We use dimensional regularization, modified to preserve susy and be applicable to solitons, and suitable renormalization conditions. 
In the mode expansion of bosonic and fermionic quantum fields,
zero modes appear then as massless nonzero modes.
\end{abstract}





\body
\section{Introduction} 

In the beginning of the 1970's particle physicists became interested in solitons. Since Dirac's work on the quantization of the electromagnetic field in the late 1920's, particles had been associated with the Fourier modes of the second-quantized fields, and perturbation theory had been used to compute scattering amplitudes. However, for the strong interactions this approach could not be used because the coupling constant is larger than unity, and nonlinearities are essential. Thus particle physicists turned to solitons as representations of particles in strongly interacting field theories. This changed the emphasis from properties of scattering amplitudes
of two or more solitons to properties of single solitons \cite{Rajaraman:1982is,Rebbi:1985wg}. 

Also in the early 1970's,  the renormalizability of nonabelian gauge theory was proven, and supersymmetry (susy)  was discovered. A natural question that arose was: are nonabelian gauge theories (and abelian gauge theories) with solitons also renormalizable? In susy theories some divergences cancel, so it seemed interesting to extend the theories with solitons to susy theories with the same solitons, and to study whether cancellations of radiative corrections did occur. In particular, the mass of a soliton gets corrections from the sum over zero-point energies of bosons and fermions. A formal proof had been constructed that in susy theories the sum of all zero-point energies cancels\cite{Zumino:1974bg}, and it was conjectured that also the corrections to the mass of a soliton vanish in susy theories\cite{D'Adda:1978mu}. We shall see that this is an oversimplification, and that the mass of solitons already receives corrections at the one-loop level.

In addition to susy, also topology became a major area of interest in soliton physics. In 1973 Nielsen and Olesen \cite{Nielsen:1973cs} used the vortex solution, which is a soliton in $2+1$ dimensions based on the abelian Higgs model with a complex scalar field, to construct topologically stable extended particles. Ginzberg and Landau \cite{Ginzburg:1950sr} had used this model to describe superconductivity in 1950, and Abrikosov \cite{Abrikosov:1956sx} had found the vortex solution in 1957. Nielsen and Olesen embedded this vortex solution into $3+1$ dimensions, and obtained in this way stringlike excitation of the dual resonance model of particle physics with a magnetic field confined inside the tubes. 't Hooft wondered if their construction could be extended to non-abelian Higgs models, and in 1974 he \cite{'tHooft:1974qc} and Polyakov \cite{Polyakov:1974ek} discovered that the nonabelian Higgs model in $3+1$ dimensions with gauge group SU(2) and a real triplet of Higgs scalars contains monopoles, which are solitonic solutions with a magnetic charge. They contain a topological number, the winding number, which prevents them from decaying to the trivial vacuum. Similarly, the vortex solution in $2+1$ dimensions has a winding number, and even the kink (a soliton in 1+1 dimensions\cite{Dashen:1975hd,Faddeev:1978rm,Gervais:1975dc}) is topologically stable. There exist also nontopological solitons\cite{Lee:1991ax} but we shall not discuss them. In 1975 Julia and Zee constructed dyons\cite{Julia:1975ff}, solitons in the SU(2) nonabelian Higgs model with an electric and a magnetic charge, and soon afterwards Prasad and Sommerfeld \cite{Prasad:1975kr} found exact expressions for these solitons in the limit of vanishing $\lambda\varphi^4$ coupling constant (the PS limit). In 1976 Bogomolnyi \cite{Bogomolny:1976de} showed that in all these cases of topological solitons one can write the energy density as a sum of squares plus total derivatives. Requiring these squares to vanish leads to first-order differential equations for solitons, the Bogomolnyi equations, which are much easier to solve than the second-order field equations. He also noted that the total energy has a bound $H\geq|Z|$ where $Z$ is the contribution from the total derivatives. This bound is called the BPS bound because for monopoles in the nonabelian Higgs model it can only be saturated for vanishing coupling constant $\lambda$. For a classical soliton at rest, $H$ is equal to its mass $M$, and $M=|Z|$. Finally in 1978 Olive and Witten \cite{Witten:1978mh} noted that the total derivative terms in the Bogomolnyi expression for the energy density are the central charges of the susy algebra of the corresponding susy theories. These charges are Heisenberg operators, containing all perturbative and nonperturbative quantum corrections. By using results of the representation theory of superalgebras in terms of physical states,  they proved that for topological solitons the BPS bound $M\geq|Z|$ must remain saturated at the quantum level: $M=|Z|$.

We shall calculate the one-loop corrections to $M$ and $Z$, and show that they are indeed equal, but nonvanishing for susy kinks\cite{Nastase:1998sy,Graham:1998qq,Shifman:1998zy,Rebhan:2002yw} and the $N=2$ monopole\cite{Rebhan:2004vn}. These calculation are not meant as a check of the proof of Olive and Witten, but rather they are a test of whether our understanding of quantum field theory in the presence of solitons has progressed enough to obtain saturation of the BPS bound. As we shall see, this is a nontrivial issue. The vacuum expectation values of the Higgs scalars acquire local corrections in the presence of solitons, boundary terms contribute to $M$ and $Z$, a new anomaly-like contribution to $Z$ yields a finite correction, and composite operators require infinite renormalization to obtain a finite answer in the ``finite" $N=4$ susy model.  We shall use the background field formalism to formulate background-covariant $R_{\xi}$ gauges, and we shall use the extended Atiyah-Singer-Patodi\cite{Atiyah:1975jf,Callias:1977kg,Weinberg:1979ma,Weinberg:1981eu} index theorem for noncompact spaces to calculate the sum over zero-point energies in the presence of solitons. We shall also introduce an extension of dimensional regularization which preserves susy and can be used for solitons\cite{Rebhan:2002uk}.\footnote{Dimensional regularization in the context of (bosonic) solitons was employed before in
Refs.~\cite{Luscher:1982wf,Parnachev:2000fz}.}

As we have discussed, solitons were initially proposed for 
describing hadrons, but when duality between electric and magnetic fields, and extended dualities in supersymmetric field theories, were developed, another point of view
emerged. It was conjectured by 
Montonen and Olive\cite{Montonen:1977sn}, and Witten\cite{Witten:1979ey} that there exist dual formulations of field theories in which particles become solitons, and solitons become particles. 
Modern work in string theory has confirmed 
and extended this hypothesis in an amazing way.

\section{The simplest case: the susy kink and its ``new anomaly"} 

In order to test one's understanding of a quantum field theory,
static quantities should be among the first to consider. In the following
we shall consider two static quantities in one of the simplest quantum
field theories with a soliton: the mass and the central charge
of the susy kink at the one-loop level. This exercise has proved
to be a surprisingly subtle topic with all kinds of pitfalls.
Even when the same renormalization conditions were employed,
different regularization methods led to
contradictory 
results\cite{Schonfeld:1979hg,Kaul:1983yt,Yamagishi:1984zv,Imbimbo:1984nq,Uchiyama:1986gf,Casahorran:1989vd,Boya:1990fp},
and this confusing state of matters lasted until the end of the
1990's, when the question was reopened by a work by two of 
us\cite{Rebhan:1997iv}, in which
it was shown that the methods used to produce the
most widely accepted result of zero corrections
in the susy case were inconsistent with the known integrability of
the bosonic sine-Gordon model \cite{Dashen:1975hd}.
Subsequently, the pitfalls of the various methods were
sorted out\cite{Nastase:1998sy,Graham:1998qq,Shifman:1998zy,Litvintsev:2000is,Goldhaber:2000ab,Goldhaber:2001rp,Goldhaber:2002mx,Bordag:2002dg,Rebhan:2002uk,Rebhan:2002yw},
which involved the discovery of an anomalous contribution to the central
charge guaranteeing BPS saturation.
In the following, we shall show how all this works out using
dimensional regularization adapted to susy solitons.

\subsection{Mass}

The mass of a soliton is obtained by taking the expectation value of the Hamiltonian with respect to the ground state in the soliton sector\footnote{%
This state is often called the soliton vacuum, but this is a misnomer because vacua have by definition vanishing energy while the soliton has a nonvanishing mass. The vacuum is the state with vanishing energy in the sector without winding, but to avoid misunderstanding, we shall consistently call it the trivial vacuum.}. In addition one needs the contribution from counter terms which are needed to renormalize the model. 

As Hamiltonian we take the gravitational Hamiltonian (obtained by varying the action with respect to an external gravitational field). We write all fields $\varphi(x,t)$ as a sum of (static) background fields $\varphi_b(x)$ and quantum fields $\eta(x,t)$, and only retain all terms quadratic in quantum fields. For real bosonic fields the Hamiltonian density of the quantum fields is of the form 
\eq
\mathcal{H}=\frac{1}{2}\dot{\eta}\dot{\eta}+\frac{1}{2}\partial_x \eta\partial_x \eta+\cdot\cdot\cdot
\eqe   
and using partial integration yields $\frac{1}{2}\partial_x \eta\partial_x \eta+\cdot\cdot\cdot=\partial_x(\frac{1}{2} \eta\partial_x \eta)-\frac{1}{2} \eta(\partial^2_x \eta+\cdot\cdot\cdot)$. The terms $-\frac{1}{2}\eta(\partial^2_x \eta+\cdot\cdot\cdot)$ are then equal to $-\frac{1}{2}\eta\ddot{\eta}$ if one uses the linearized field equations for the fluctuations, and the expectation value $-\frac{1}{2}\langle soliton|  \eta\ddot{\eta} |soliton\rangle$ is equal to $+\frac{1}{2}\langle soliton|  \dot{\eta}\dot{\eta} |soliton\rangle$. For a real (Majorana) fermion there are no background fields, and the Hamiltonian density is of the form $\mathcal{H}=\2\bar{\psi}\g^1\partial_x\psi+\ldots$ . Again using the linearized field equations for the fermion's fluctuations, one finds $\mathcal{H}=\frac{i}2\psi^{\dagger}\dot{\psi}$ since $\bar{\psi}=\psi^{\dagger}i\g^0$ and $(\g^0)^2=-1$. Thus the one-loop quantum corrections to the mass of a soliton, $M=\langle soliton|  \int\mathcal{H}dx |soliton\rangle$, are of the generic form
\eqa
\nonumber M^{(1)}=&\int&[\langle \dot{\eta}\dot{\eta}\rangle+\frac{i}2\langle\psi^{\dagger}\dot{\psi}\rangle]dx\\
\nonumber &+&{\rm boundary\;\;terms}\;\;\int\partial_x(\frac{1}{2}\langle \eta\partial_x\eta\rangle)\\
&+&{\rm counter\;\;terms}\;\;\Delta M\,.
\eqae 

To define the infinite and finite parts of the one-loop corrections, we need a regularization scheme that preserves susy and is easy to work with. This singles out dimensional regularization. Usually one needs dimensional regularization by dimensional reduction to preserve susy, but that option is not available to us because the soliton occupies all space dimensions. Going up in dimensions in general violates susy, but there is a way around these objections which combines the virtues of both approaches. In all cases we consider, the susy action in $D+1$ dimensions can be rewritten as a susy action in $D+2$ dimensions. Then going down in dimensions, we use in $(D+\epsilon)+1$ dimensions standard dimensional regularization. This scheme clearly preserves susy, and it leaves enough space for the soliton. 

Let us see how things work out for the susy kink. The susy action (after eliminating the susy auxiliary field) is given by 
\eqa
\nonumber\mathcal{L}&=&\frac{1}{2}\dot{\varphi}^2-\frac{1}{2}(\partial_x\varphi)^2-\frac{1}{2}U^2-\frac{1}{2}\bar{\psi}\hspace{2pt}\slash\hspace{-6pt}\partial\psi-\frac{1}{2}U'\bar{\psi}\psi,\\
&&\frac{1}{2}U^2=\frac{\lambda}{4}(\varphi^2-\mu_{0}^2/\lambda)^2,
\eqae
where $\psi$ is a 2-component Majorana spinor and $\varphi$ a real scalar field. This model has $N=(1,1)$ susy in 1+1 dimensions, but the same expression for $\mathcal{L}$ can also be viewed as an $N=1$ model in 2+1 dimensions. The operator $\7{\partial}$ in the Dirac action and in the transformation law $\delta\psi=(\7{\partial}\varphi-U)\epsilon$  is then given by $\g^0\partial_0\varphi+\g^1\partial_x\varphi+\g^2\partial_y\varphi$. 

The energy density obtained from the gravitational stress tensor reads 
\eqa
\mathcal{H}=\frac{1}{2}\dot{\varphi}^2+\frac{1}{2}(\partial_k\varphi)^2+\frac{1}{2}U^2+\frac{1}{2}\bar{\psi}\g^k\partial_k\psi+\frac{1}{2}U'\bar{\psi}\psi\;,\;\;k=1,2.
\eqae
For the classical soliton solution we set $\dot\varphi=\psi=0$ and denote $\varphi_b$ by $\varphi_K$.
The classical mass of the kink follows from the Bogomolnyi way of writing the classical Hamiltonian as a sum of squares plus a boundary term 
\eqa
\nonumber\mathcal{H}&=&\frac{1}{2}(\partial_x\varphi_K)^2+\frac{1}{2}U_K^2\\
&=&\frac{1}{2}(\partial_x\varphi_K+U_K)^2-(\partial_x\varphi_K)U_K,
\label{total}
\eqae
where $U_K=U(\varphi_K)$. Thus the classical field equation for the soliton reads $\partial_x\varphi_K+U_K=0$, and the classical mass is 
\eq
M_{cl}=-\int_{-\infty}^{+\infty}dx\,\partial_x[\int_0^{\varphi_K(x)}U(\varphi')d\varphi']=\frac{2\sqrt{2}\mu_{0}^3}{3\lambda}\,.
\eqe

The kink solution is given by $\varphi_K(x)=\frac{\mu}{\sqrt{\lambda}}\tanh\frac{\mu x}{\sqrt{2}}$ (where $\mu$ is the normalized mass introduced below), but we shall not need this.
Decomposing $\varphi$ into $\varphi_K(x)+\eta(x,y,t)$ , we find for the terms quadratic in quantum fluctuations
\eqa
\mathcal{H}^{(2)}=\frac{1}{2}\dot{\eta}\dot{\eta}+\frac{1}{2}(\partial_k \eta)(\partial_k \eta)+\frac{1}{2}(\frac{1}{2}U_K^{2})''\eta\eta+\frac{1}{2}\bar{\psi}\g^k\partial_k\psi+\frac{1}{2}U_K'\bar{\psi}\psi\,,\;\;
\eqae  
where $k=1,2$. Partial integration, use of the linearized field equations for quantum fields $\eta$ and $\psi$, and substitution of $\bar{\psi}\g^0=-i\psi^{\dagger}$ yields
\eqa
\nonumber\mathcal{H}^{(2)}=\frac{1}{2}\dot{\eta}\dot{\eta}+\frac{1}{2}\partial_k(\eta\partial_k\eta)-\frac{1}{2}\eta\ddot{\eta}+\frac{i}{2}\psi^{\dagger}\dot{\psi}\\
\langle\mathcal{H}^{(2)}\rangle=\langle\dot{\eta}\dot{\eta}+\frac{1}{2}\partial_k(\eta\partial_k\eta)+\frac{i}{2}\psi^{\dagger}\dot{\psi}\rangle.
\label{A}
\eqae
(We shall later choose a real (Majorana) representation for the Dirac matrices, and then $\psi$ is real, thus $\psi^\dagger=\psi^T$.)

To renormalize the field theory with quantum fields $\eta$ and $\psi$, one considers the trivial vacuum, and chooses as background field $\varphi_b=\frac{\mu}{\sqrt{\lambda}}$. There are terms with 2, 3, and 4 quantum fields, and the terms with three $\eta$'s,
or one $\eta$ and two $\psi$'s, 
can give a tadpole loop which is divergent and needs renormalization. We therefore decompose the bare mass $\mu^2_0$ into a renormalized part $\mu^2$ and a counter term $\Delta\mu^2$, and require that $\Delta\mu^2$ cancels all (finite as well as infinite) contributions  of the tadpoles. The bosonic loop yields $\Delta\mu^2 =3\lambda\langle \eta^2\rangle$ while the fermionic loop yields  $\Delta\mu^2 =-2\lambda\langle \eta^2\rangle$ 
\eq
\nonumber\vcenter{\hbox{\includegraphics[scale=0.5]{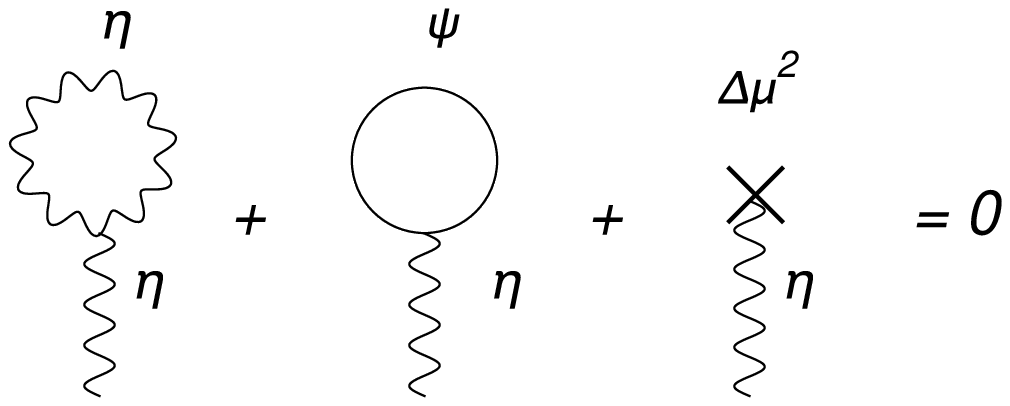}\raisebox{24pt}{,}}}\;\;\;\begin{array}{c}\displaystyle{\mu^2_0=\mu^2+\Delta\mu^2,} \\\quad\\ \Delta\mu^2=\displaystyle{\lambda\int\frac{d^{2+\epsilon}k}{(2\pi)^{2+\epsilon}}\int\frac{-i}{k^2+m^2-i\epsilon}}\\
\displaystyle{\quad\;=\lambda\int\frac{d^{1+\epsilon}k}{(2\pi)^{1+\epsilon}}\frac{1}{2\sqrt{k^2+m^2}}\,.\phantom{\Bigg|^1}
}
\end{array}
\eqe
No further renormalizations are needed, so the $Z$ factors for $\lambda$, $\eta$ and $\psi$ are all unity. This is a particular set of renormalization conditions. 

Having fixed $\Delta\mu^2$ in the trivial sector, we now return to the kink sector and find for the mass counter term at the one-loop level 
\eq
\Delta M=(\frac{2\sqrt{2}}{3\lambda})(\mu^2+\Delta\mu^2)^{\frac{3}{2}}-M_{cl}=\frac{m\Delta\mu^2}{\lambda}\;;\;m=\sqrt{2}\mu\,.
\eqe
We must now evaluate the terms in (\ref{A}). We do this by expanding $\eta$ and $\psi$ into modes, but as we shall see, the sum $\langle\dot{\eta}\dot{\eta}+i\psi^{\dagger}\dot{\psi}\rangle$ can also be extracted from an index theorem. 

The field equation for $\eta(x,y,t)=\phi(x)e^{ily}e^{-i\omega t}$ reads 
\eq
-\partial^2_x\phi+(\frac{1}{2}U_K^2)''\phi=(\omega^2-l^2)\phi\,.
\eqe
Actually, the field operator $-\6_x^2+(\frac{1}{2}U_K^2)''$ factorizes into $(-\partial_x+m\tanh\frac{mx}{2})(\partial_x+m\tanh\frac{mx}{2})\equiv L^{\dagger}_2 L_2$, and this allows explicit expressions for the zero mode $\phi_0 (x)$ (satisfying $[-\partial^2_x+(\frac{1}{2}U_K^2)'']\phi_0=0$, with $\omega^2_0=0$), the bound state $\phi_B (x)$ (with $\omega^2_B=-\frac{3}{4}m^2$), and the continuous spectrum 
$\phi(k,x)$ (with $\omega^2_k=k^2+m^2$). However, we do not need explicit expressions for these functions. The mode expansion in 1+$\epsilon$ spatial dimensions reads 
\eqa\label{kinkmodeexp}
\nonumber\eta(x,y,t)&=&\int_{-\infty}^\infty\frac{d^{\epsilon}l}{(2\pi)^{\frac{\epsilon}{2}}}\biggl\{\int_{-\infty}^\infty\frac{dk}{\sqrt{2\pi}}\frac{1}{\sqrt{2\omega_{kl}}}(a_{kl}\phi(k,x)e^{ily}e^{-i\omega_{kl}t}\\
\nonumber&&\hspace{4cm}
+a^{\dagger}_{kl}\phi^{*}(k,x)e^{-ily}e^{i\omega_{kl}t})\\
\nonumber&&+\frac{1}{\sqrt{2\omega_{Bl}}}(a_{Bl}\phi_{B}(x)e^{ily}e^{-i\omega_{Bl}t}+a^{\dagger}_{Bl}\phi_{B}(x)e^{-ily}e^{i\omega_{Bl}t})\\
&&+\frac{1}{\sqrt{2\omega_{0l}}}(a_{0l}\phi_{0}(x)e^{ily}e^{-i|l|t}+a^{\dagger}_{0l}\phi_{0}(x)e^{-ily}e^{i|l|t})\biggr\},
\eqae 
where $\omega^2_{kl}=k^2+l^2+m^2$, $\omega^2_{Bl}=-\frac{3}{4}m^2+l^2$, and $\omega^2_{0l}=l^2$. 
The annihilation and creation operators $a$ and $a^\dagger$ satisfy the usual commutation relations, for example $[a_{0l},a_{0l'}^\dagger]=\delta^\epsilon(l-l')$.
The functions $\phi_B(x)$ and $\phi_0(x)$ are normalized to unity, while the distorted plane waves $\phi(k,x)$ are normalized such that they become plain waves $e^{i(kx+\frac{1}{2}\delta(k))}$ for $x\rightarrow+\infty$ and $e^{i(kx-\frac{1}{2}\delta(k))}$ for $x\rightarrow-\infty$, satisfying the completeness relation\footnote{The completeness relation\cite{Goldhaber:2001rp} reads $\int\phi(k,x)\phi^{*}(k,x')\frac{dk}{2\pi}+\phi_B(x)\phi_B(x')+\phi_0(x)\phi_0(x')=\delta(x-x')$, and has been rewritten in terms of $(\phi(k,x)\phi^{*}(k,x')-e^{ik(x-x')})$ by bringing the delta function to the left-hand side. It follows that $\phi(k,x)$, $\phi_{B}(x)$ and $\phi_0(x)$ are orthonormal, for example $\int\phi(k,x)\phi^{*}(k',x)dx=2\pi\delta(k-k')$.}
\eq
\int_{-\infty}^{\infty}(|\phi(k,x)|^2-1)\frac{dk}{2\pi}+\phi^2_B(x)+\phi^2_0(x)=0.
\eqe
  
For the fermion we use a real (Majorana) representation of the Dirac matrices $\g^{\mu}$ which diagonalizes the iterated field equations in 2+1 dimensions
\eq
\g^1=\left(\begin{array}{cc}1 & 0 \\0 & -1\end{array}\right)\;,\;\g^0=\left(\begin{array}{cc}0 & -1 \\1 & 0\end{array}\right)\;,\;\g^2=\left(\begin{array}{cc}0 & 1 \\1 & 0\end{array}\right)
\label{rep}
\eqe
and also makes $\psi$ real.
The field equation
\eq
(\hspace{2pt}\slash\hspace{-6pt}\partial+U_K')\psi=0\;;\;U_K'=m\tanh\frac{mx}{2}
\eqe
reads then in component form
\eq
\left.\begin{array}{c}(\partial_x+U_K')\psi_+=(\partial_0-\partial_y)\psi_{-} \\(\partial_x-U_K')\psi_-=(\partial_0+\partial_y)\psi_{+}\end{array}\right\}\;\;\psi=\left(\begin{array}{c}\psi_{+} \\ \psi_{-}\end{array}\right)       
\eqe
The iterated field equation of $\psi_+$ is the same as the $\eta$ field equation, while for $\psi_{-}$ we find the conjugate field operator
\eq
\left.\begin{array}{c}(L^{\dagger}_2 L_2-\partial^2_y+\partial_0^2)(\eta\; {\rm or}\; \psi_+)=0 \\(L_2 L_2^{\dagger}-\partial^2_y+\partial_0^2)\psi_{-}=0\end{array}\right\}\;\;\begin{array}{c}L_2=\partial_x+U_K' \\ L_2^{\dagger}=-\partial_x+U_K'\end{array}       
\eqe
Setting $\psi_{\pm}=\psi_{\pm}(x)e^{ily-i\omega t}$, the Dirac equation yields 
\eq
\psi_{-}(k,x)=\frac{i(\partial_x+U'_K)}{\omega_{kl}+l}\psi_+(k,x)\;;\;\omega^2_{kl}=k^2+l^2+m^2
\eqe
The mode expansion of $\psi$ in 1+$\epsilon$ spatial dimensions 
is then given by
\eqa
\nonumber\psi=\left(\begin{array}{cc}\psi_{+} \\ \psi_{-}\end{array}\right)&=&\int_{-\infty}^{\infty}\frac{d^{\epsilon}l}{(2\pi)^{\frac{\epsilon}{2}}}\int_{-\infty}^{\infty}\frac{dk}{\sqrt{2\pi}}\biggl\{\\
\nonumber&&\frac{1}{\sqrt{2\omega_{kl}}}\biggl[b_{kl}\left(\begin{array}{cc}\sqrt{\omega_{kl}+l}\phi(k,x) \\ \sqrt{\omega_{kl}-l}is(k,x)\end{array}\right)e^{ily}e^{-i\omega_{kl} t}\\ \nonumber&&\qquad\qquad+b^{\dagger}_{kl}\left(\begin{array}{cc}\sqrt{\omega_{kl}+l}\phi(k,x)^{*} \\ \sqrt{\omega_{kl}-l}(-i)s(k,x)^{*}\end{array}\right)e^{-ily}e^{i\omega_{kl} t}\biggr]\\
\nonumber &+&\frac{1}{\sqrt{2\omega_{Bl}}}\biggl[b_{Bl}\left(\begin{array}{cc}\sqrt{\omega_{Bl}+l}\phi_{B}(x) \\ \sqrt{\omega_{Bl}-l}is_{B}(x)\end{array}\right)e^{ily}e^{-i\omega_{Bl} t}\\ \nonumber&&\qquad\qquad+b^{\dagger}_{Bl}\left(\begin{array}{cc}\sqrt{\omega_{Bl}+l}\phi_{B}(x) \\ \sqrt{\omega_{Bl}-l}(-i)s_{B}(x)\end{array}\right)e^{-ily}e^{i\omega_{Bl} t}\biggr]\\
\nonumber&+&\frac{1}{\sqrt{2|l|}}\biggl[b_{0l}\left(\begin{array}{cc}\sqrt{|l|+l}\,\phi_{0}(x) \\ 0\end{array}\right)e^{ily}e^{-i|l| t}\\ \nonumber&&\qquad\qquad+b^{\dagger}_{0l}\left(\begin{array}{cc}\sqrt{|l|+l}\,\phi_{0}(x) \\ 0\end{array}\right)e^{-ily}e^{i|l| t}\biggr]\biggr\}\\
\label{crazy}
\eqae 
where 
\eq\label{sphi}
 s(k,x)=\frac{(\partial_{x}+U'_K)\phi(k,x)}{\omega_{k}}\;,\;\;\omega^2_{k}=k^{2}+m^{2}.
\eqe
Several remarks are to be made 
\begin{itemize}
  \item we have extracted the same factors $\frac{1}{\sqrt{2\omega}}$ as for the boson;
  \item the normalization factors $\sqrt{\omega+l}$ and $\sqrt{\omega-l}$ are needed to satisfy the equal-time canonical anticommutation relations, as we shall check,
  \eqa
  \nonumber\{\psi_{\pm}(x,y,t),\psi_{\pm}(x',y',t)\}&=&\delta(x-x')\delta^{\epsilon}(y-y'),\\
\{\psi_{+}(x,y,t),\psi_{-}(x',y',t)\}&=&0
\label{lhs}
  \eqae 
  \item we treat zero modes and nonzero modes on equal footing. In fact, the zero modes have become massless nonzero modes at the regularized level
with energy $|l|$;
  \item there are no zero modes for $\psi_{-}$, while the zero modes of $\psi_+$ have only positive momenta $l$ in the extra dimensions, yielding massless chiral domain-wall fermions, which are right-moving on the domain wall;
  \item 
the zero mode sector can also be written as 
  \eq
  \int_{-\infty}^\infty\frac{dl}{(2\pi)^{\frac{\epsilon}{2}}}b_{0l}\phi_0(x)e^{il(y-t)}
  \eqe
  where for positive $l$, $b_{0l}$ is an annihilation operator, but for negative $l$ a creation operator ($b_{0,-l}=b^{\dagger}_{0,l}$);
  \item the normalization factor $\sqrt{\omega-l}$ for $s_k$ in $\psi_{-}$ is obtained as follows: given that $\psi_{+}(k,x)$ is written in terms of $\sqrt{\omega+l}\phi$, multiply $\psi_{-}(k,x)=\frac{i(\partial_x+U'_K)}{\omega+l}\psi_+(k,x)$ in the numerator and denominator by $\sqrt{\omega-l}$
  \eqa
\sqrt{\omega+l}\psi_{-}(k,x)&=&\sqrt{\omega+l}\frac{\sqrt{\omega-l}}{\sqrt{\omega-l}}\frac{i(\partial_x+U'_K)}{\omega+l}\phi(k,x)\\
&=&\sqrt{\omega-l}\frac{i(\partial_x+U'_K)}{\sqrt{\omega^2-l^2}}\phi(k,x)=\sqrt{\omega-l}is(k,x);  \nonumber
  \eqae   
\item the reality of $\psi$ is manifest. One can also write the spinors in the
terms with $e^{i\omega t}$ as $\sqrt{\omega+l}\phi$ and
$-\sqrt{\omega-l}is$
since $\phi(k,x)^*=\phi(-k,x)$ and thus also $s(k,x)^*=s(-k,x)$, which corresponds to
$\psi=-C\bar \psi^T$ where $C=i\gamma^0$ is the charge conjugation matrix.
(The relation $\phi(k,x)^*=\phi(-k,x)$ follows from the reflection symmetry
$x\to-x$ of the action, but one can also read it off from the explicit
expression for $\phi(k,x)$.\cite{Goldhaber:2001rp})
\end{itemize}
 
Let us check that this mode expansion for $\psi_{\pm}$ is correct by assuming that the annihilation and creation operators satisfy the usual anticommutators, and verifying that we obtain $\delta(x-x')\delta^{\epsilon}(y-y')$ and zero in (\ref{lhs}). We begin with 
\eqa
&&\{\psi_+(x,y,t),\psi_{+}(x',y',t)\}=\int\frac{dk}{2\pi}\int\frac{d^{\epsilon}l}{(2\pi)^{\epsilon}}\\
\nonumber&&\left[\frac{\omega_{kl}+l}{2\omega_{kl}}\{\phi(k,x)\phi^{*}(k,x')e^{il(y-y')}+\phi^{*}(k,x)\phi(k,x')e^{-il(y-y')}\}\right.\\
\nonumber&&+\left.\{\frac{\omega_{Bl}+l}{2\omega_{Bl}}\phi_B(x)\phi_B(x')+\frac{|l|+l}{2|l|}\phi_0(x)\phi_0(x')\}(e^{il(y-y')}+e^{-il(y-y')})\right]\,.
\eqae
Using $\phi^{*}(k,x)=\phi(-k,x)$, and changing the integration variable for the terms with $\phi^{*}(k,x)$ from $k$ to $-k$, we find that all terms factorize into terms with $\omega+l$ times $e^{il(y-y')}+e^{-il(y-y')}$. The factors $l$ in $\omega+l$ cancel by symmetric integration, and then also the terms with $\omega$ cancel. All terms are now proportional to $e^{il(y-y')}$, and integration over $l$ yields the required $\delta^{\epsilon}(y-y')$. One is left with 
\eq
\int\phi(k,x)\phi^{*}(k,x')\frac{dk}{2\pi}+\phi_B(x)\phi_B(x')+\phi_0(x)\phi_0(x')
\eqe
which is indeed equal to $\delta(x-x')$.

For the $\{\psi_-,\psi_-\}$ anticommutator there are two differences: instead of $\phi(k,x)$ one has $s(k,x)$, and there are no zero modes. One finds 
\eq
\int s(k,x)s^{*}(k,x')\frac{dk}{2\pi}+s_B(x)s_B(x').
\eqe
This is again equal to $\delta(x-x')$, as it is the completeness relation for $L_2 L^{\dagger}_2$. One can also directly check this\footnote{
Use Eqs.~(9) and (10) of Ref.~\cite{Goldhaber:2001rp} together with Eq.~(\ref{sphi}) above.}. For the $\{\psi_-,\psi_+\}$ anticommutator one finds along the same lines 
\eqa
&&\{\psi_+(x,y,t),\psi_{-}(x',y',t)\}=\int\frac{dk}{2\pi}\int\frac{d^{\epsilon}l}{(2\pi)^{\epsilon}}\\
\nonumber&&\left[\frac{\sqrt{\omega_{kl}^2-l^2}}{2\omega_{kl}}\left\{\phi(k,x)(-i)s^{*}(k,x')e^{il(y-y')}+is(k,x)\phi^{*}(k,x')e^{-il(y-y')}\right\}\right.\\
\nonumber&&+\left.\frac{\sqrt{\omega_{Bl}^2-l^2}}{2\omega_{Bl}}\left\{\phi_B(x)(-i)s_B(x')e^{il(y-y')}+is_B(x)\phi_B(x')e^{il(y-y')}\right\}\right]\,.
\eqae
Because there are now no terms linear in $l$ which multiply the exponents $e^{il(y-y')}$, we can change the integration variables $k$ and $l$ to $-k$ and $-l$ in half of the terms, and, using $\phi(k,x)^*=\phi(-k,x)$ and $s(k,x)^*=s(-k,x)$, all terms cancel.  

The calculation of the one-loop mass of the susy kink is now simple. We must evaluate 
\eq\label{M1contributions}
M^{(1)}=\int dx \int d^{\epsilon} y\langle \dot{\eta}\dot{\eta}+\frac{i}{2} \psi^{T}\dot{\psi}\rangle+\left.\int d^{\epsilon} y\frac{1}{2}\langle \eta\partial_x\eta\rangle\right|^{x=\infty}_{x=-\infty}+\frac{m}{\lambda}\Delta\mu^2
\eqe 
The first term gives the sum over zero-point energies 
\eqa\label{sum0pt}
\int dxd^\e y\; \langle \dot{\eta}\dot{\eta}+\frac{i}{2} \psi^{T}\dot{\psi}\rangle&=&V_y\int dx\int\frac{dk}{2\pi}\int\frac{d^{\epsilon}l}{(2\pi)^{\epsilon}}\\
\nonumber&&\hspace{-3.5cm}\times\frac{\omega_{kl}}{2}\biggl[\phi^{*}(k,x)\phi(k,x)-\frac{\omega_{kl}+l}{2\omega_{kl}}\phi^{*}(k,x)\phi(k,x)
-\frac{\omega_{kl}-l}{2\omega_{kl}}s^{*}(k,x)s(k,x)\biggr]\\
\nonumber
&=&V_y\int dx\int\frac{dk}{2\pi}\int\!\frac{d^{\epsilon}l}{(2\pi)^{\epsilon}}
\frac{\omega_{kl}}{4}(|\phi(k,x)|^2-|s(k,x)|^2)
\eqae
where $V_y$ is the volume $\int d^{\epsilon}y$ of the extra dimensions
and where only contributions from the continuous spectrum have remained.
There is no contribution from the bound state because $\int dx(\varphi^2_B(x)-s^2_B(x))$ vanishes (partially integrate as in (\ref{time}), there is no boundary term because $\varphi_B(x)$ falls off exponentially fast). There is also no contribution from the zero mode because the corresponding integral $\int dk d^{\epsilon} l \;l^2/|l|$ is a scaleless integral, and scaleless integrals vanish in dimensional regularization. 
Note that the terms proportional to a single power of $l$ (arising from the $\sqrt{\w+l}$ and $\sqrt{\w-l}$ in (\ref{crazy})) drop out because they are odd in the loop momentum $l$; in the calculation of $Z$ these terms will give a crucial contribution. 
The total derivative $\int dx\frac{\partial}{\partial x}\int d^{\e}y \langle\eta\partial_x\eta\rangle$ does not contribute because  $\eta\partial_x\eta=\frac{1}{2}\partial_x(\eta\eta)$, and $\langle \eta\eta\rangle$ can only depend on $x$ as $\frac{1}{x}$, in which case the derivative $\partial_k$ yields $\frac{1}{x^2}$ which vanishes for large $x$.\footnote{Actually, $\langle \eta\eta\rangle$ falls off even faster then $1/x$, namely exponentially fast.\cite{Shifman:1998zy,Goldhaber:2001rp}}(In $3+1$ dimensions one can get a contribution because there the measure is $4\pi r^2$).  

The expression in (\ref{sum0pt}) is what 
in early approaches was believed to be zero, but
which is actually infinite. Combining it with the counter term contribution
in (\ref{M1contributions}),
the total mass per volume $V_y$ becomes then 
\eq
M^{(1)}=\int_{-\infty}^{\infty}\frac{dk}{2\pi}\int\frac{d^{\e}l}{(2\pi)^{\e}}\frac{\omega_{kl}}{4}\Delta\rho(k^2)+\frac{m}{\lambda}\Delta\mu^2
\eqe
where 
\eq
\Delta\rho(k^2)=\int_{-\infty}^{\infty}dx(|\phi(k,x)|^2-|s(k,x)|^2)
\eqe
is the difference of spectral densities of $\psi_+$ and $\psi_-$. One can use an index theorem\cite{Atiyah:1975jf,Callias:1977kg,Weinberg:1979ma,Weinberg:1981eu,Rebhan:2006fg} to compute $\Delta\rho(k^2)$, or one can directly calculate it, using partial integration,
\eqa\label{Deltarhocalc}
\nonumber\int |s(k,x)|^2 dx&=&\int \frac{[(\partial_x+U')\phi^{*}(k,x)][(\partial_x+U')\phi(k,x)]}{\w_k^2}dx\\
\nonumber&=&\left.[\frac{\phi^{*}(k,x)(\partial_x+U')\phi(k,x)}{\w^2_k}]\right|_{x=-\infty}^{x=\infty}\\
\nonumber&&\qquad+\int_{-\infty}^{\infty}\frac{\phi^{*}(k,x)(-\partial_x+U')(\partial_x+U')\phi(k,x)}{\omega_k^2}dx\\
&=&\frac{2m}{k^2+m^2}+\int_{-\infty}^{\infty}dx|\phi(k,x)|^2\,.
\label{time}
\eqae
We used that
since $\phi^{*}(k,x)\partial_x\phi(k,x)=ik$ and $U'\rightarrow\pm m$ as $x\rightarrow\pm\infty$, the terms with $\phi^{*}(k,x)\partial_x\phi(k,x)$ cancel, while the terms with $U'$ add. Note that $\Delta\rho(k^2)$
is nonvanishing, because $|\phi(k,x)|^2$ of the continuous spectrum 
in the second line of (\ref{Deltarhocalc})
does not vanish as $x\to\pm\infty$.
With this result for the difference of spectral densities we obtain
\eqa
\nonumber M^{(1)}&=&\frac{m}{2}\int \frac{dk d^\e l}{(2\pi)^{1+\e}}\left[ -\frac{\sqrt{k^2+l^2+m^2}}{k^2+m^2}+\frac{1}{\sqrt{k^2+l^2+m^2}}\right]\\
&=&\frac{m}{2}\int \frac{dk d^\e l}{(2\pi)^{1+\e}}\left[\frac{-l^2}{(k^2+m^2)\sqrt{k^2+l^2+m^2}}\right].
\eqae
Note that the extra dimensions, needed to maintain susy at the regularized
level, have produced a nonvanishing correction proportional to the square
of the momentum in the extra dimensions!
Using the standard formula for dimensional regularization 
\eq
\int\frac{d^n l}{(l^2+ \mathcal{M} ^2)^\alpha}=\pi^{n/2}(\mathcal{M} ^2)^{\frac{n}{2}-\alpha}\frac{\Gamma(\alpha-\frac{n}{2})}{\Gamma(\alpha)},
\eqe
we find for the $l$ integral 
\eqa
\nonumber\int\frac{d^\epsilon l\;l^2}{(l^2+ \mathcal{M} ^2)^{\frac{1}{2}}}&=&\int\frac{d^\epsilon l}{(l^2+ \mathcal{M} ^2)^{-\frac{1}{2}}}-\mathcal{M} ^2\int\frac{d^\epsilon l}{(l^2+ \mathcal{M} ^2)^{\frac{1}{2}}}\\
&=&\pi^{\frac{\epsilon}{2}}(\mathcal{M}^2)^{\frac{\epsilon}{2}+\frac{1}{2}}(\frac{-\epsilon}{\epsilon+1})\frac{\Gamma(\frac{1}{2}-\frac{\epsilon}{2})}{\Gamma(\frac{1}{2})}\,,
\eqae 
where $\mathcal M^2=k^2+m^2$.
We are left with the $k$ integral
\eq
\int dk(k^2+m^2)^{\frac{\epsilon}{2}-\frac{1}{2}}=-\frac{2}{\epsilon}\quad{\rm for}\;\epsilon\rightarrow0\,.
\eqe
The factors $\epsilon$ and $\frac{1}{\epsilon}$ cancel, and the final result is
 \eq
 M^{(1)}=-\frac{m}{2\pi}\,.
 \eqe

\subsection{Central Charge} 

The central charge is one of the generators of the susy algebra. To construct the latter, we begin with the Noether current for rigid susy. If one integrates its time-component over space, one obtains the susy charge $Q$, but it is advantageous to postpone this integration and first evaluate the susy variation of the susy current, $\delta j^{\mu}=-i[j^{\mu},\bar{Q}\e]$.\cite{Shifman:1998zy} Extracting $\e$, and integrating over space yields the $\{Q,Q\}$ anticommutators. 

In order to regularize the quantum corrections, we first construct the $\{Q,Q\}$ anticommutators in $2+1$ dimensions, and then descend to $(1+\e)+1$ dimensions. In $1+\e$ dimensions, the translation generators $P_y$ in the direction of the $\e$ extra dimensions are still present, and they are added to the central charge $Z_x$ which one naively finds in $1+1$ dimensions. As we shall show, $-P_y+Z_x$ is the regularized central charge. In loop calculations $P_y$ will give a finite but nonvanishing contribution. For bosons in the loop, symmetric integration over $l$ gives a vanishing result, but for fermions in the loop, a factor $l$ coming from the derivative $\frac{\partial}{\partial y}$ in $P_y$ combines with another factor $l$ coming from the normalization factors $\sqrt{\w+l}$ and $\sqrt{\w-l}$ of the spinors $\psi_+$ and $\psi_-$ to give a factor $l^2$. Integration over $l$ yields then a nonvanishing contribution,
\eq
\langle P_y\rangle\sim\int\langle\psi_+\partial_y\psi_+ +\psi_-\partial_y\psi_-\rangle\sim\int\frac{l^2}{\w}\Delta\rho(k^2)\neq0\,.
\label{consistency}
\eqe    
This result has the same form  as one encounters in the calculation of the chiral triangle anomaly using dimensional regularization, namely a factor $l^2$ in the numerator which yields a factor $n$ as $n\rightarrow 0$, and a divergent loop integral which gives a factor $\frac{1}{n}$. We therefore refer to the term in (\ref{consistency}) as an anomaly-like contribution, or, less precisely, as an anomaly. (There is no anomaly in the conservation of the central charge current, just as there is no anomaly in the ordinary susy current, or the stress tensor, but there is an anomaly in the conservation of the ${\it conformal}$ current \cite{Rebhan:2002yw}). The final result for the one-loop contributions to the regularized central charge is equal to the one-loop mass correction, and thus BPS saturation continues to hold at the quantum level. 

In earlier work, not enough attention was paid to careful regularization, and extra terms, such as the occurrence of $P_y$, were missed. Any other regularization scheme should also lead to BPS saturation if one is careful enough. Of course, one should specify the same renormalization conditions in the calculation of $M^{(1)}$ and $Z^{(1)}$; in our case this means that we again remove tadpoles by decomposing $\mu_0^2$ into $\mu^2+\Delta\mu^2$. Let us now show the details for the kink. 

The Noether current (in $2+1$ dimensions) is given by $j^{\mu}=-\hspace{2pt}\slash\hspace{-6pt}\partial\varphi\g^{\mu}\psi-U\g^{\mu}\psi$, and with the representation in (\ref{rep}) we find for the two spinor components 
\eqa\label{j0+}
\nonumber j^{0}_{+}=(\dot{\varphi}-\partial_y\varphi)\psi_+ + (\partial_x+U)\psi_-\\
j^{0}_{-}=(\dot{\varphi}+\partial_y\varphi)\psi_- + (\partial_x-U)\psi_+
\eqae
We can evaluate the variation of $j^0_\pm$ either by transforming the fields in $j^0_\pm$ under rigid susy transformations, or by evaluating the anticommutators with $Q_\pm=\int j^0_\pm(x',y',t)dx'dy'$. We follow the latter approach. Using the equal-time canonical (anti)commutation relations
\eqa
\nonumber[\dot{\varphi}(x',y',t),\varphi(x,y,t)]&=&\frac{1}{i}\delta(x'-x)\delta(y'-y)\\
\nonumber\{\psi_\pm(x',y',t),\psi_\pm(x,y,t)\}&=&\delta(x'-x)\delta(y'-y)\\
\{\psi_+(x',y',t),\psi_-(x,y,t)\}&=&0
\eqae 
one finds straightforwardly,  after partial integration of $\frac{\partial}{\partial x'}$ and $\frac{\partial}{\partial y'}$  derivatives, 
\eqa
\nonumber\{Q_+,j_+\}&=&\dot{\varphi}^2-2\dot{\varphi}\partial_y\varphi+(\partial_y\varphi)^2+(\partial_x\varphi)^2 +2U\partial_x\varphi+U^2\\
&&-2i\psi_+\partial_y\psi_+-2iU\psi_+\psi_-+i\psi_+\partial_x\psi_-+i\psi_-\partial_x\psi_+
\eqae
The right-hand side can be written in terms of the densities of the Hamiltonian, translation generator $P_y$, and naive central charge $Z_x$ as follows
\eqa
\nonumber\{Q_+,j_+\}&=&2\mathcal{H}-2\mathcal{P}_y+2\mathcal{Z}_x\\
\nonumber2\mathcal{H}&=&\dot{\varphi}^2+(\partial_y\varphi)^2+(\partial_x\varphi)^2 +U^2-2iU\psi_+\psi_-\\
\nonumber&&+i\psi_+\partial_x\psi_-+i\psi_-\partial_x\psi_+ -i\psi_+\partial_y\psi_+ +i\psi_-\partial_y\psi_- \\
\nonumber-2\mathcal{P}_y&=&-2\dot{\varphi}\partial_y\varphi-i\psi_+\partial_y\psi_+ -i\psi_-\partial_y\psi_-\\
2\mathcal{Z}_x&=&2U\partial_x\varphi
\eqae
The sum of the last term of $2\mathcal{H}$ and $-2\mathcal{P}_y$ cancels, but we have added these terms to obtain the complete expressions for $\mathcal{H}$ and $\mathcal{P}_y$. One can check that $\mathcal{H}$ and $\mathcal{P}_y$ generate the correct time- and space- translations of $\varphi$, $\dot{\varphi}$, $\psi_+$, and $\psi_-$. 
The other susy anticommutators are given by
\eqa
\nonumber\{Q_-,j_-\}&=&2\mathcal{H}+2\mathcal{P}_y-2\mathcal{Z}_x\\
\{Q_+,j_-\}&=&2\mathcal{P}_x+2\mathcal{Z}_y\\
\eqae
where $2\mathcal{Z}_y=2U\partial_y\varphi$.

Integrating over $x$ and $y$, and
using two-component spinors we obtain
\be
\2 \{ Q,Q \} = - (\gamma^\mu \gamma^0) P_\mu + (\gamma^2\gamma^0)Z_x-
(\gamma^1\gamma^0)Z_y
\ee
where $P_0=H$, and this clearly demonstrates that $Z_x-P_y$ and $P_x+Z_y$
are the regulated versions of $Z_x$ and $P_x$, respectively.

The naive central charge $Z_x$ receives no quantum corrections. This was observed by several authors. To demonstrate this, we expand $\varphi=\varphi_K+\eta$ and $\mu_0^2=\mu^2+\Delta\mu^2$, and find to second order in $\eta$
\eqa
\nonumber\mathcal{Z}_x&=&U\partial_x\varphi=\partial_x(\int^{\varphi}U(\varphi')d\varphi')\\
&=&U_K\partial_x\varphi_K+\partial_x(U_K\eta)+\frac{1}{2}\partial_x(U'_K\eta^2)-\frac{\Delta\mu^2}{\sqrt{2\lambda}}\partial_x\varphi_K
\eqae 
The first term yields classical BPS saturation, since it is just minus the total derivative in (\ref{total}). Taking the expectation value in the kink ground state, the term linear in $\eta$ vanishes, and the last two terms give, after integration over $x$ and $y$, 
\eq
\langle Z^{(1)}_x\rangle=\left[m\langle\eta^2(x\rightarrow\infty)\rangle-2\Delta\mu^2\frac{\mu}{\sqrt{2}\lambda}\right]V_y.
\eqe 
We used that $U'_K\to\pm m$ 
and $\varphi_K\to\pm \mu/\sqrt\lambda$
as $x\rightarrow\pm\infty$. 
Recalling that $\mu=m/\sqrt{2}$, and $\Delta\mu^2=\lambda\langle\eta^2\rangle$ in the trivial vacuum, we see that $\langle Z^{(1)}_x\rangle$ vanishes. The tadpole renormalization in the trivial vacuum, and thus also far away from the kink, cancels the contribution from the naive central charge. 

However, we get a nonvanishing correction from $P_y$. The bosonic fluctuation do not contribute
\eq
\langle P_y^{bos}\rangle=\int\langle \dot{\eta}\partial_y\eta\rangle dxd^\e y\sim\int\frac{\w l}{2\w}|\phi(k,x)|^2dkd^\e l=0
\eqe
due to symmetric integration over $l$. But from the fermions we get a nonvanishing contribution
\eqa\label{Pyferm}
\nonumber\langle P_y^{ferm}\rangle&=&\int\frac{i}{2}\langle \psi_+\partial_y\psi_+ +\psi_-\partial_y\psi_-\rangle dxd^\e y\\
\nonumber&=&\frac{1}{2}\int\frac{dk}{2\pi}\frac{d^\e l}{(2\pi)^\e}(\frac{l(\w+l)|\varphi(k,x)|^2}{2\w}+\frac{l(\w-l)|s(k,x)|^2}{2\w})dxd^\e y\\
&=&\frac{1}{2}\int\frac{dk}{2\pi}\frac{d^\e l}{(2\pi)^\e}\frac{l^2}{2\w}(|\varphi(k,x)|^2-|s(k,x)|^2)dxd^\e y
\eqae
This is the same expression as we found for $M^{(1)}$, hence BPS saturation holds. 

Repeating the same calculation for $N=2$ susy $\varphi^4$ kinks, one
finds\cite{Nastase:1998sy,Rebhan:2002uk} that BPS saturation holds without anomalous contributions from
$\langle P_y \rangle$, because in these models the extra
fields lead to a complete cancellation of $\Delta\rho(k^2)$. 
However, in the 1+1-dimensional 
$N=2$ CP$^1$ model with so-called twisted mass term\cite{Dorey:1998yh}, $\Delta\rho(k^2)$
is instead twice the amount found in the minimally susy kink models.%
\footnote{Another special feature of the $N=2$ twisted-mass
CP$^1$ model is that the nonrenormalization of $\langle Z_x \rangle$
involves fermionic boundary terms.\cite{Mayrhofer:2007ms}}
The appearance of an anomalous contribution in the $N=2$ twisted-mass
CP$^1$ model\cite{Shifman:2006bs,Mayrhofer:2007ms} has
to do with the fact that the $N=2$ CP$^1$ model provides
an effective field theory for confined monopoles\cite{Hanany:2003hp,Auzzi:2003fs,Auzzi:2003em,Shifman:2003uh} of 3+1-dimensional 
$N=2$ SU(2)$\times$U(1) gauge theories, which in turn are
related\cite{Shifman:2004dr} 
by holomorphicity to 't Hooft-Polyakov $N=2$ monopoles,
and for the latter we shall indeed find anomalous
contributions to the central charge in what follows.
   
\section{Boundary terms and composite operator
renormalization for supersymmetric monopoles} 

We now discuss susy monopoles in 3+1 dimensions, and study
how BPS saturation is realized when one-loop quantum corrections
are included. From what has been learned from the kink,
one might expect that if one defines proper renormalization
conditions and takes again into account an anomaly-like contribution
to the central charge, BPS saturation will follow.
This turns out to be the case for the $N=2$ monopole
and leads us to correcting once again previous results in the
literature\footnote{In contrast to the susy kink, the few explicit
calculations of one-loop corrections to the $N=2$ monopole
were all agreeing on a null result
in a minimal renormalization scheme\cite{Kaul:1984bp,Imbimbo:1985mt}.},
but, surprisingly enough, for the $N=4$ monopole in the
``finite'' $N=4$ super Yang-Mills theory, there are divergences
left in boundary contributions, and these can only be canceled,
it seems, by introducing a new concept in the study of solitons,
which was not necessary before: infinite composite operator
renormalization of the stress tensor and the central charge current. For the $N=2$ model, all surface contributions, 
which are individually divergent, cancel nicely.

Composite operator renormalization of
the stress tensor and the central charge is no
contradiction to the lore that ``conserved currents don't renormalize'', because that applies only to internal currents, not to
spacetime ones. The stress tensor appearing in the susy algebra
can be written as the sum of
an improved stress tensor, which is traceless, and ``improvement
terms'' corresponding to $R\varphi^2$ terms in the action in curved space. While the improved stress tensor turns out to
be finite, the non-traceless part renormalizes multiplicatively
in the $N=4$ model, and just happens to be finite as well in the $N=2$ case.
Thus this new feature of composite operator renormalization in the $N=4$
model does not upset the BPS saturation of the $N=2$ model
that was obtained previously without it.
One could of course start with the improved currents at the
classical level, but this would change the traditional value
of the classical mass of the 't Hooft-Polyakov monopole.

\subsection{The $N=2$ monopole}

The action of the $N=2$ super Yang-Mills model in 3+1 dimensions
can be obtained in a simple way by applying dimensional reduction
to the action of minimal
super Yang-Mills theory in 5+1 dimensions
\begin{eqnarray}
  \mathcal L&=&-\frac{1}{4} F_{MN}^2-\bar{\lambda}\Gamma^M D_M\lambda;
\qquad \lambda={\psi\choose 0}\nonumber\\
&=&-
  \frac{1}{4} F_{\mu\nu}^2-\frac{1}{2}(D_\mu P)^2
-\frac{1}{2}(D_\mu S)^2-\frac{1}{2} 
  g^2(S\times P)^2\nonumber\\
  \label{eq:L4d} 
  &&-
  \bar\psi\gamma^\mu D_\mu\psi 
-g\bar\psi\gamma_5(P\times\psi)-ig\bar\psi (S\times \psi).
\end{eqnarray}
with $\psi$ a 4-component complex spinor and $(S\times P)^a=\epsilon^{abc}
S^bP^c$.
We decomposed $A_M^a$ into $(A_\mu^a,P^a,S^a)$ and used a particular representation of the Dirac matrices in $5+1$ dimensions \cite{Rebhan:2004vn}. In the topologically
trivial sector we take $S^3$ as
the Higgs field with vev $v$ (and $S^1$, $S^2$ the would-be Goldstone fields).
In the soliton sector, the energy density for a static configuration
with nonvanishing $A_j^a$ and $S^a$ can be written as
\be
\mathcal H=\4\left(F_{ij}^a+\epsilon_{ijk}D_k S^a\right)^2
-\2\6_k\left(\epsilon_{ijk}F_{ij}^a S^a\right).
\ee
Thus the Bogomolnyi equation for a monopole residing in
$A_j$ and $S$ reads
\begin{equation}
F_{ij}^a+\epsilon_{ijk}D_k S^a=0.
\end{equation}
The asymptotic behavior of $A_j$ and $S$ is given by
\bea
\nonumber
A_j^a&=&\epsilon_{aij}\frac{x^j}{gr^2}+\ldots,
\quad F_{ij}^a=-\epsilon_{ijk}\frac{x^ax^k}{gr^4}+\ldots, \\
S^a&=&\frac{x^a v}{r}-\frac{x^a}{gr^2}+\ldots,
\quad D_k S^a =\frac{x^ax^k}{gr^4}+\ldots,
\label{monopoleasympt}
\eea
where the suppressed subleading terms are exponentially decreasing
for large radius $r$, and the classical mass of the monopole
reads
\be
M_{cl}=\frac{4\pi v}{g}=\frac{4\pi m}{g^2}
\ee
with $m=gv$.

The susy algebra can be obtained as before by varying the time component
of the Noether current and afterwards integrating over space. One obtains\footnote{ For $\{Q^{\alpha},Q^{\beta}\}$ one finds the integral of a total derivative of a bilinear in fermions, $\int\partial_j(\psi^T C\gamma^0\psi)d^3x(\gamma^jC^{-1})^{\alpha\beta}$. Since $\langle\psi\psi\rangle$ vanishes, we shall omit this term from the algebra.}
\be
\2\{ Q^\alpha, Q^\dagger_\beta \}
=\delta^\alpha_\beta P_0
-(\gamma^k \gamma^0)^\alpha{}_\beta P_k-
(\gamma_5 \gamma^0)^\alpha{}_\beta U+i(\gamma^0)^\alpha{}_\beta V
\label{qq}
\ee 
where
\bea
\label{UV}
U&=&\int d^3x\,  \6_k \left[\2\epsilon^{ijk} F_{ij}\cdot S
+F_{k0}\cdot P \right],\nn
V&=&\int d^3x\, \6_k\left[\2\epsilon^{ijk} F_{ij}\cdot P
- F_{k0}\cdot S \right],
\eea
and $P_\mu=\int T_{\mu0}\,d^3x$ so that $P_0=H>0$.
To make contact with the usual form of the susy algebra for
$N$-extended susy,
\bea
\2\{ Q^{Ai}, Q^{Bj} \} &=& \epsilon^{AB} Z^{ij},\quad
Z^{ij}=-Z^{ij}\;{\rm complex} \nonumber\\
\2\{ Q^{Ai}, \bar Q_{\dot B j} \} &=& \delta^i_j (\sigma^\mu)^A{}_{\dot B}
P_\mu\,,
\eea
note that our complex $Q^\alpha$ can be written in terms of
Majorana $Q^{j\alpha}$ as $Q^\alpha=(Q^1+iQ^2)^\alpha$,
and $Q^{\alpha j}=(Q^{Aj},\bar Q_{\dot A j})$
in terms of two-component spinors. Then $Z^{12}=-Z^{21}
=-U+iV$ for the $N=2$ model, whereas we already see that for the
$N=4$ model to be discussed below there will be 6 complex
(12 real) central charges. Classically only $U$ is nonvanishing,
and BPS saturation holds for the above monopole solution.\footnote{ Using a suitable representation of the Dirac matrices, the right-hand side of (\ref{qq}) takes on the form $\left(\begin{array}{cc}P_0+\sigma^k P_k & iU+V \\-iU+V & P_0-\sigma^k P_k\end{array}\right)$. For vanishing $P_k$ one obtains $P_0^2\geq U^2+V^2$, hence in general, $M^2\geq U^2+V^2$}

For calculating quantum corrections, we use an ``$R_\xi$'' gauge-fixing
term
\be
\mathcal L_{\rm fix}=-\frac1{2\xi} \left(D_M(A)a^M\right)^2
=-\frac1{2\xi} \left(
D_\mu(A)a^\mu+g P\times p+g S\times s \right)^2.
\ee
We have written ``$R_\xi$'' in quotation marks because
a genuine $R_\xi$ gauge-fixing term would have a factor $\xi$
in front of $g P\times p$ and $g S\times s$. The above form
is advantageous to keep the SO(5,1) symmetry of the theory
prior to dimensional reduction. We shall set $\xi=1$
in which case the kinetic terms in the fluctuation equations
become diagonal (in a genuine $R_\xi$ gauge, this is also
true for $\xi\not=1$).

The field equations for the fluctuations $a_{\9 m}=\{a_i,s\}$,
$i=1,2,3$, read
\be
 \left((\6_0^2-\6_5^2-D_{\9\ell}^2)\delta_{\9 m\9 n}
  -2 g {F}_{\9 m \9 n} \times \right)a_{\9 n}=0,
\ee
where $D_{\9\ell}^{ab}=(D_i^{ab},igS^{ab})$ with
$D_i^{ab}=\6_i\delta^{ab}+\epsilon^{acb}A_\mu^c$
and $S^{ab}=\epsilon^{acb}S^c$.
They can be written in spinor notation as
\be\label{DbarD}
(\5{\7{\sf D}}{\7{\sf D}}+\6_5^2-\6_0^2)\5{\7a} = 0;\quad
\bar{\7{\sf D}}{\7{\sf D}}=
    D^2_{\9 m}+\2 \bar\sigma^{\9 m \9 n}g{F}_{\9 m\9 n},
\ee
where $\bar{\7a}=\bar\sigma^{\9 m}a_{\9 m}$,
$\bar\sigma^{{\underline m}{\underline n}}=\2(\bar\sigma^{\underline m}
\sigma^{\underline n}-\bar\sigma^{\underline n}\sigma^{\underline m})$
with $\bar\sigma^{\underline m}=(\vec \sigma,-i\mathbf 1)$ and
$\sigma^{\underline m}=(\vec \sigma,i\mathbf 1)$ in the
4-dimensional Euclidean space labeled by the index $\underline m$.
Furthermore,
\be\label{DDbar}
({\7{\sf D}}\5{\7{\sf D}} +\6_5^2-\6_0^2)\7q=0; \quad
{\7{\sf D}}\bar{\7{\sf D}}=D^2_{\9 m}
\ee
for the remaining quartet of bosonic fields
$q_{\9 m}=(a_0,p,b,c)$, where $b,c$ are Faddeev-Popov ghost fields.

For the spinors we find
\begin{equation}
  \label{eq:f1}
  {\7{\sf D}} \psi_+ = (\6_0-\6_5)\psi_-\quad ,\quad 
  \bar{\7{\sf D}} \psi_- = (\6_0+\6_5)\psi_+\ ,
\end{equation}
where
\be
{\7{\sf D}}^{ab}=\sigma^{\9m}D_{\9m}^{ab}=\sigma^k D_k^{ab}+igS^{ab},\quad
\5{\7{\sf D}}^{ab}=\5\sigma^{\9m}D_{\9m}^{ab}=\sigma^k D_k^{ab}-igS^{ab}.
\ee
Iterating (\ref{eq:f1}) we have
\be
\label{fermeoms}
\bar{\7{\sf D}}{\7{\sf D}}\,\psi_+=(\6_0^2-\6_5^2)\psi_+\,,\quad  
{\7{\sf D}}\bar{\7{\sf D}}\,\psi_-=(\6_0^2-\6_5^2)\psi_-\,,
\ee
so the two columns of $\5{\7a}$ have the same field equations as
$\psi_+$, and the two columns of $\7q$ have the same field equations
as $\psi_-$, the analogous situation as we found for the susy kink.

One can now construct the gravitational stress tensor $T_{\mu\nu}$
and consider the terms in the Hamiltonian density $T_{00}$ which are
quadratic in quantum fields. For the bosons, there are terms of the
form $\6a\6a$ and terms of the form $a\6^2a$. Partially integrating
the former, we can use the field equations for the fluctuations to
obtain the following result\cite{Rebhan:2006fg}
\bea
M^{\rm 1-loop}&=&\int d^3x \langle
a_0\6_0^2a_0-a_{j}\6_0^2a_{j} - p\6_0^2p -s\6_0^2s 
-b\6_0^2c-(\6_0^2b)c\nonumber\\
&&\qquad+{i\02}\psi^\dagger
\olrp_0 \psi\rangle\nonumber\\
&&
+\lim_{r\to\infty}\4 4\pi r^2 \frac{\6}{\6r}
\langle
a_0^2+a_j^2+p^2+s^2+2bc-\frac23 a_j^2 
\rangle,
\eea
where we used that $\6_j\6_k\langle a_j a_k\rangle=\frac13 \6_k^2 
\langle a_j^2 \rangle$ and $\langle a_j a_0 \rangle = 0$ for
large $r$.

The bulk contributions give the sum over zero-point energies of all
quantum fields. Fermions have the mode expansion
\bea
\label{eq:fqf}
\psi\!&=&\!
{ \psi_+ \choose i\psi_- } =
\int\frac{d^{\epsilon}\ell}{(2\pi)^{\epsilon/2}}
\int{d^3k\0(2\pi)^{3/2}}{1\0\sqrt{2\omega}}\sum
\biggl\{ b_{kl} e^{-i(\omega t - \ell y)}
{\sqrt{\omega+\ell}\; \chi_k^+ \choose - \sqrt{\omega-\ell} \;\chi_k^- }\nonumber\\
&& +d_{kl}^\dagger e^{i (\omega t - \ell y)}
{ \sqrt{\omega+\ell} \;\chi_k^+ \choose  \sqrt{\omega-\ell} \;\chi_k^- }
\biggr\}\ +\mathrm{bound\ states + zero\  modes}\ ,
\eea
where the sum refers to the two possible polarizations of the $\chi_k$'s.
On the other hand, the mode expansion of the bosonic fields
$a_j$ and $s$ (which we
combined into $\7a$) only involves $\chi_k^+$ and the one
of the quartet $\7q$ only $\chi_k^-$.
This leads to
\be\label{E0pt}
M^{(1)\rm bulk}=
V_y\int d^3x \int {d^3k\,d^\epsilon\ell \0 (2\pi)^{3+\epsilon}}
{\omega\02}(\mathcal N_+|\chi^+_k|^2(x)+\mathcal N_-|\chi^-_k|^2(x))
\ee
where $\mathcal N_+=4-2$ from $a_{\9m}$ and $\psi_+$, and
$\mathcal N_-=1+1-1-1-2$ from $q_{\9m}=(a_0,p,b,c)$ and $\psi_-$.
The result thus involves only a difference of spectral densities
which can be evaluated by an index theorem\cite{Weinberg:1979ma,Weinberg:1981eu,Kaul:1984bp,Imbimbo:1985mt,Rebhan:2006fg}
\be
\label{drh}
\Delta\rho(k^2)=\int d^3x (|\chi^+_k|^2(x)-|\chi^-_k|^2(x))
=\frac{-4\pi m}{k^2(k^2+m^2)}.
\ee

On the other hand, all surface contributions (in the present $N=2$
case) cancel,\cite{Rebhan:2006fg}
\bea\label{Msurfcancel}
M^{(1)\rm surface}&=&\lim_{r\to\infty}{1\04}4\pi r^2
{\6\0\6r}\langle a_0^2+a_j^2+p^2+s^2+2bc-\frac23 a_j^2\rangle\nn
&=&(-1+3+1+1-2-2)\lim_{r\to\infty}\pi r^2{\6\0\6r}\langle s^2\rangle=0,
\eea
where we have used that the propagators of all bosonic fields become
the same for large $r$, since only terms of order $1/r$ can contribute,
whereas $F_{\mu\nu}^a$ falls off as $1/r^2$.
(The contribution of $\langle a_0^2\rangle$ is minus $\langle s^2\rangle$
because of the metric $\eta^{\mu\nu}$ in the canonical commutation
relations of the creating and annihilation operators.)
Hence, $M^{(1)\rm bulk}$ is the complete, but still
unrenormalized, one-loop result.

The momentum integral that we are left with to evaluate upon
insertion of (\ref{drh}) into (\ref{E0pt}) is UV divergent,
and the required counter term $\Delta M$ comes from the
renormalization of $M_{\rm cl.}=4\pi v_0/g_0$. We clearly need
$Z_g$ and $Z_v$. In the background field formalism which we
have been using, one has the well-known relation $Z_g=Z_A^{-1/2}$,
so we could first determine $Z_A$ by requiring that all loops
with two external background fields $A_\mu^3$ are cancelled at
zero external momentum (the quantum fields in these loops are
all massive, so there are no IR problems). We did this in
Ref.~\cite{Rebhan:2006fg} even at arbitary $\xi$, but one can
get $Z_g$ also from the known one-loop formula of the $\beta$-function
\bea
Z_g&=&1-g^2\{ \frac{11}3-\frac23n_{\rm Maj.ferm.}-\frac16n_{\rm
real\; scalars} \} C_2({\rm SU(2)}) \frac{I}2\nn
&=&1-2g^2I\quad \mbox{for}\;N=2;\quad
\mbox{but $Z_g=1$ for $N=4$},
\eea
where
\be\label{Idef}
I\equiv\int {d^{4+\epsilon}k\0(2\pi)^{4+\epsilon}}
{-i\0 (k^2 + m^2)^2 }=
\int {d^{4+\epsilon}k_E\0(2\pi)^{4+\epsilon}}
{1\0 (k^2_E + m^2)^2 } = -{1\08\pi^2}{1\0\epsilon}+O(\epsilon^0).
\ee

The value of $Z_v$ is equal to $Z_S$, because constant $v$'s are a special
case of arbitrary background fields $S^3$ (more precisely, in the trivial sector only the combination $v+S^3$ occurs). At $\xi=1$, the value
of $Z_S$ is equal to $Z_A$ because all relevant background-field
vertices are contained in $a_M D_m^2 a^M$ which is SO(5,1) invariant.
(At $\xi\not=1$, $Z_S$ becomes $\xi$-dependent, while $Z_A$ is
$\xi$-independent, because it is given by the $\beta$-function.)
Since $Z_g Z_S^{1/2}=1$, the mass $m=gv$ does not renormalize (at $\xi=1$),
and thus
\be
\Delta M={4\pi m\0Z_g^2 g^2}-{4\pi m\0g^2}
={4\pi m\0g^2}4g^2 I=16\pi m I.
\ee

The mass correction to the $N=2$ monopole is finally given by
\bea\label{E0ptmon}
M^{(1)}&=&2
\int d^3x \int {d^3k\,d^\epsilon\ell \0 (2\pi)^{3+\epsilon}}
{\sqrt{k^2+\ell^2+m^2}\02}\Delta\rho(k)+\Delta M\nn
&=&-2{m\0\pi} {\Gamma(-\2-{\epsilon\02}) \0 (2\pi^\2)^\epsilon \Gamma(-\2)}
\int_0^\infty dk (k^2+m^2)^{-\2+{\epsilon\02}}+16\pi m I\nn
&=&\left(-{1\01+\epsilon}+1\right)16\pi mI=-{2m\0\pi}+O(\epsilon).
\eea

The one-loop corrections to the original expression in (\ref{UV}) for the
central charge $U$ of the $N=2$ monopole
cancel completely\footnote{The first graph above (\ref{graph}) yields a divergence $-4g^2 I U$, but wave function renormalization of $S$ and $A_{\mu}$ in U yields a counterterm $4g^2 I U$. This cancellation was worked out first
in Ref.~\cite{Imbimbo:1985mt}, but it only
works for $N=2$, while $N=4$ involves new issues\cite{Rebhan:2005yi,Rebhan:2006fg}
that we shall discuss
below.}
against the counterterms due to ordinary renormalization
, but the translation
operator $P_y$ in the extra $\epsilon$ dimensions gives again
a nonvanishing ``anomalous'' contribution which exactly matches $M^{(1)}$,
in complete analogy to the case of the susy kink (see Eq.~\ref{Pyferm})%
\footnote{However, in contrast to the case of the susy kink, if
one combines the integral with $\Delta\rho$ in the mass
correction (\ref{E0ptmon}) with the integral representation
of the counter term $\Delta M$, one does not obtain a
factor $\ell^2$ in the numerator as in (\ref{eq:an2}).},
\be\label{eq:an2}
U^{(1)}=P_y=
\int {d^3k\,d^\epsilon\ell \0 (2\pi)^{3+\epsilon}}
{\ell^2 \0 2 \sqrt{k^2+\ell^2+m^2}}\,\Delta\rho(k^2) 
=-{2m\0\pi}+O(\epsilon).
\ee
Clearly, BPS saturation holds for the $N=2$ monopole at the one-loop
level. However, the finite nonvanishing correction to both the mass and
the central charge had been missed in all the literature preceding
Ref.~\cite{Rebhan:2004vn}, although closer inspection reveals that
the commonly accepted
null result was in conflict with the low-energy effective action
of $N=2$ super Yang-Mills theory
obtained some time ago by Seiberg and 
Witten\cite{Seiberg:1994rs,Seiberg:1994aj}.

\subsection{The $N=4$ monopole}

We now turn to the monopole in $N=4$ super Yang-Mills
theory in 3+1 dimensions, 
where the naive expectation of vanishing one-loop corrections
to mass and central charge in the end turns out to be correct.
However, how this comes about is highly nontrivial, and in
several ways the properties of the $N=4$ case are opposite
to the $N=2$ case, with dramatic consequences.

We begin by following the same steps as in the $N=2$ case.
The action of $N=4$ super Yang-Mills
theory in 3+1 dimensions is most easily obtained by applying
dimensional reduction to the $N=1$ super Yang-Mills theory
in 9+1 dimensions, yielding
\bea
  \label{eq:L10d}
    \mathcal L
&=&- \4 F_{MN}^2 - \2 \bar{\lambda}\Gamma^M D_M\lambda\\
&=&-\4 F_{\mu\nu}^2-\2(D_\mu S_{\3j})^2-\2(D_\mu P_{\3j})^2
-\2 \5\lambda^I \7D \lambda^I + \mbox{interactions}.\nonumber
\eea
where we decomposed 
$A_M^a$ into $(A_\mu^a,S_{\3j}^a,P_{\3j}^a)$, with 3 adjoint scalars
$S_{\3j}$ and 3 pseudoscalars $P_{\3j}$, ${\3j}=1,2,3$, instead of only one of
each
in the $N=2$ case. The 16-component adjoint Majorana-Weyl spinor $\lambda^a$
has been decomposed into four 4-component Majorana spinors $\lambda^{aI}$
with $I=1,\ldots,4$, with a factor $\2$ in front of their action
because of the Majorana property. The susy algebra reads
\bea\label{QQ}
\2\{ Q^{\alpha I},Q^{\beta J} \} &=&
\delta^{IJ} (\gamma^\mu C^{-1})^{\alpha\beta} P_\mu \nn
&&+i(\gamma_5 C^{-1})^{\alpha\beta} (\alpha_{\3j})^{IJ} \intx U_{\3j}
-(C^{-1})^{\alpha\beta} (\beta_{\3j})^{IJ} \intx V_{\3j} \nn
&&+(C^{-1})^{\alpha\beta} (\alpha_{\3j})^{IJ} \intx \tilde V_{\3j}
+i(\gamma_5 C^{-1})^{\alpha\beta} (\beta_{\3j})^{IJ} \intx \tilde U_{\3j}\nn
&&-\frac18 \intx (\5\lambda \Gamma^0 \Gamma_{PQ}D_R\lambda)(\Gamma^{PQR}C^{-1})^{\alpha\beta},
\eea
where the last term is on-shell a total derivative\footnote{%
Use $\Gamma^{RS}\Gamma^N D_N\lambda=0=\Gamma^{RSN}D_N\lambda
+\Gamma^R D^S \lambda - \Gamma^S D^R\lambda$ to write all terms as $\bar{\lambda}\Gamma^{RST}\lambda$. Then use 
$\5\lambda\Gamma^{RST}D_N\lambda=\2\6_N(\5\lambda\Gamma^{RST}\lambda)$.}
of the form $\6_\rho(\5\lambda \Gamma^{0RS}\lambda)$.
Since the expectation value of this term contains the spinor
trace tr$(\Gamma^{0RS}\7k)$, which vanishes, we drop this term from now on.
In (\ref{QQ}) we have used a particular\cite{Rebhan:2004vn} representation
of the 32$\times$32 Dirac matrices $\Gamma^M$ in terms of the 4$\times$4
Dirac matrices, involving the matrices $\alpha_{\3j}$ and $\beta_{\3j}$ 
which are proportional to the matrices $\eta_{\3j}^{IJ}$ and 
$\5\eta_{\3j}^{IJ}$ which 't Hooft introduced for the construction
of instantons. The $\alpha_{\3j}$ and $\beta_{\3j}$ respresent
the 6 generators of SO(4): totally antisymmetric 4$\times$4 matrices,
purely imaginary, and either self-dual ($\alpha$) or anti-self-dual ($\beta$).
The indices $I$ and $J$ are raised and lowered with the Euclidean
metric $\delta^{IJ}$ and $\delta_{IJ}$, and finally $\alpha,\beta=1,\ldots,4$
are the spinor indices in 3+1 dimensions. Clearly, we have 12 real
central charges
\bea\label{UUVV}
&&U_{\3j}=\6_i(S^a_{\3j} \2 \epsilon^{ijk} F_{jk}^a),
\qquad \tilde U_{\3j}=\6_i(P^a_{\3j} F_{0i}^a) \nn
&&V_{\3j}=\6_i(P^a_{\3j} \2 \epsilon^{ijk} F_{jk}^a),
\qquad \tilde V_{\3j}=\6_i(S^a_{\3j} F_{0i}^a).
\eea
In the $N=2$ case, Eqs.~(\ref{UV}),
only the sums $U_3+\tilde U_3$ and $V_3+\tilde V_3$ appeared,
but here they split into parts with different tensor structures,
half of them with $\alpha$ matrices, the other half with $\beta$'s.

We set $S_3^a=v$ for adjoint color index $a=3$ in the topologically trivial
sector, and locate the monopole inside the fields $A_j^a$ and $S_3^a$.
For quantum calculations we use again the background field formalism
as in the $N=2$ case above, which now gives
$Z_v=Z_S=Z_A=Z_g=1$, since the $\beta$-function for the $N=4$ model
vanishes.

The gravitational stress tensor yields the Hamiltonian density,
which we write again after use of the linearized field equations
for fluctuations as time derivatives giving the sum over zero-point
energies, and surface terms\cite{Rebhan:2006fg}
\bea\label{M1loopN4}
M^{\rm 1-loop}&=&\int d^3x \langle
a_0\6_0^2a_0-a_{j}\6_0^2a_j - p_{\3j}\6_0^2p_{\3j} -s_{\3j}\6_0^2s_{\3j} -b\6_0^2c-(\6_0^2b)c\nonumber\\
&&\qquad+{i\02}(\lambda^I)^T\6_0 \lambda^I
\rangle\nonumber\\
&&
+\lim_{r\to\infty}\4 4\pi r^2 \frac{\6}{\6r}
\langle
a_0^2+a_j^2+p^2_{\3j}+s^2_{\3j}+2bc-\frac23 a_j^2
\rangle,
\eea
where the only differences with the $N=2$ model are that
there are now three times as many scalar and pseudoscalar fields and we have
four real 4-component spinors instead one one complex
4-component spinor. However, the consequences could not have
been more severe. The sum of the zero-point energies 
(the bulk contribution in (\ref{M1loopN4}))
vanishes
for the $N=4$ case: the fields associated with the field
operator $\5{\7{\sf D}}{\7{\sf D}}$ are $a_j,s_3,\lambda_+^I$
and yield in eq.~(\ref{E0pt}) $\mathcal N^+=3+1-4=0$, instead of $3+1-2=2$, while the fields
associated with ${\7{\sf D}}\5{\7{\sf D}}$ are $a_0,s_1,s_2,p_{\3j},
b,c$, and $\lambda_-^I$
yielding $\mathcal N^-=1+1+1+3-1-1-4=0$, too, instead of $1+1-1-1-2=-2$.
On the other hand, for $N=4$ the boundary terms no longer vanish, since
instead of $-1+3+1+1-2-2=0$ 
we now have $-1+3+3+3-2-2=4$, yielding
\be\label{Msurf4l}
M^{(1)\rm surface}=\lim_{r\to\infty}\pi r^2{\6\0\6r}\langle 4 (s_1^a)^2\rangle.
\ee
For large $r$, the difference between
the operators $\5{\7{\sf D}}{\7{\sf D}}$ and ${\7{\sf D}}\5{\7{\sf D}}$
is due to $F_{\9m\9n}$, see (\ref{DbarD}) and (\ref{DDbar}),
which falls off like $1/r^2$, 
so the bosonic propagators in the background covariant
$\xi=1$ gauge have a common asymptotic form,
\be
\langle a_M^a(x) a_N^b(y) \rangle \simeq \eta_{MN} G^{ab}(x,y),\quad
\langle b^a(x) c^b(y) \rangle \simeq - G^{ab}(x,y),\quad
\ee
with
\be
G^{ab}(x,y)=\langle x|
{-i\0-\Box+m^2-\frac{2m}{r}}(\delta^{ab}-\hat x^a \hat x^b)
+{i\0\Box}\hat x^a \hat x^b | y \rangle.
\ee
Inserting complete sets of momentum eigenstates, the same procedure
as used to calculate anomalies from index theorems\cite{Rebhan:2006fg} yields for the
$r$-dependent part of $G^{aa}(x,x)$
\be
\langle s_1^a(x) s_1^a(x) \rangle \simeq
2\int {d^{4+\epsilon}k\0(2\pi)^{4+\epsilon}}{-i\0(k^2+m^2)+2ik^\mu\6_\mu-\6_\mu^2
-{2m\0r}} \to 2 \frac{2m}{r} I
\ee
with $I$ given in (\ref{Idef}). (The overall factor 2 is due
to tracing $\delta^{ab}-\hat x^a \hat x^b$.)

Hence, we have arrived at a divergent result for the mass of the
$N=4$ monopole,
\be\label{M1N4div}
M^{\rm 1-loop}=M^{(1)\rm surface}=-16\pi m I.
\ee
Ordinary renormalization, namely renormalization of the parameters
in the action, is of no help, since, as we have seen, all $Z$ factors
which helped to make the $N=2$ result finite, are unity in the
$N=4$ case.

The solution is extra-ordinary renormalization, namely
renormalization of the stress tensor density, and also of the
central charge density (and, in fact, all currents of the
corresponding susy multiplet) as composite operators.
In the literature it has been shown that the improved stress tensor\cite{Callan:1970ze}
does not renormalize\cite{Callan:1970ze,Freedman:1974gs,Freedman:1974ze,Collins:1976vm,Brown:1980pq}, 
which we extend to the statement that none of
the improved currents in the susy multiplet renormalize\cite{Hagiwara:1979pu,Rebhan:2005yi}.
However, the currents in the susy algebra displayed above are
nonimproved currents, and to construct improved currents one must add
improvement terms to the unimproved currents. In order to find
those for both the stress tensor and the central charges, we go back
one step and begin with our unimproved Noether susy currents $j^\mu$,
to which we add the improvement terms $\Delta j_{\rm imp}^\mu$,
\be
j^\mu_{\rm imp}=j^\mu+\Delta j_{\rm imp}^\mu
=\2 \Gamma^{RS} F_{RS}^a \Gamma^\mu \lambda^a - \frac23
\Gamma^{\mu\nu}\6_\nu (A_{\mathcal J}^a \Gamma^{\mathcal J} \lambda^a)
\ee
where we use a 10-dimensional notation in which
$A_{\mathcal J}$, $\mathcal J=5,\ldots 10$, comprises all
scalars and pseudoscalars in the model. The $\Gamma$-matrices
are 32$\times$32-matrices but we in fact deal with the dimensionally
reduced theory so that the sum over $\nu$ runs only from 0 to 3. Both $j^{\mu}$ and $j^\mu_{\rm imp}$ are on-shell conserved. (Use the Bianchi identity $\Gamma^{RS\mathcal{J}}A_{\mathcal{J}}\times F_{RS}=0$). In addition, 
the improved (ordinary, not conformal) susy currents are on-shell
gamma-traceless, 
$
\Gamma_\mu j^\mu_{\rm imp}=0.
$ 
This can be verified by using $\Gamma_\mu \Gamma^{\rho\sigma}\Gamma^\mu=0$
and $F_{\rho \mathcal J}=D_\rho A_{\mathcal J}$. One finds
$
\Gamma_\mu j^\mu_{\rm imp}=2 \Gamma^{\rho \mathcal J} (D_\rho A_{\mathcal J})
\cdot\lambda+2\Gamma^{\mathcal J \mathcal K} (A_{\mathcal J}\times
A_{\mathcal K})\cdot \lambda
-2\Gamma^\nu \6_\nu(A_{\mathcal J} \Gamma^{\mathcal J} \cdot \lambda)
$
which indeed vanishes on-shell, where $\Gamma^\rho D_\rho \lambda=
-\Gamma^{\mathcal K}A_{\mathcal K}\times\lambda$.
From the susy variation of $\Delta j_{\rm imp}^\mu$ we find the
improvement terms in $T_{\mu\nu}$ and the central charges.
Switching to 3+1-dimensional notation, we find\cite{Rebhan:2006fg}
\bea
\label{T00imprN4}
\Delta T^{\rm impr}_{00}&=&-\frac16 \6_j^2 (A_{\mathcal J} A^{\mathcal J})\\
\label{UimprN4}
\Delta U^{\rm impr}_3&=&-{1\03}\left[U_3+\int {i\08}\6_i (\epsilon^{ijk}
\bar \lambda \alpha^1 \gamma_{jk} \lambda) d^3x \right].
\eea

It is important to note that we do not start with improved Noether
currents at the classical level. Indeed, the standard result for
the classical value of the mass of a monopole is only obtained when the
unimproved stress tensor is used, and also the classical value of the improved and unimproved central charge differ.\footnote{This is clear from
the fact that $\Delta U^{\rm impr}$ involves the bosonic term
$U$.} However, even when starting with unimproved currents, we have to expect improvement terms as counterterms,
since we can write the unimproved currents as $j_\mu=j_\mu^{\rm impr}-
\Delta j_\mu^{\rm impr}$ and we expect the improved part to be finite,
and the improvement terms $\Delta j_\mu^{\rm impr}$ to renormalize
multiplicatively. Denoting the common $Z$ factor for all improvement
terms in the susy current, the stress tensor and the central charges
by $Z_{\rm impr}$, the composite operator counterterms to mass and
central charge will be given by
\be
\Delta T_{00}=-(Z_{\rm impr}-1)\Delta T^{\rm impr}_{00},\quad
\Delta U = -(Z_{\rm impr}-1) \Delta U^{\rm impr},
\ee
where the overall minus sign is due to having written 
$j_\mu=j_\mu^{\rm impr}-\Delta j_\mu^{\rm impr}$.
We shall now show by a detailed calculation\cite{Rebhan:2005yi} that a single factor 
$Z_{\rm impr}$ removes the divergences in mass and central
charge and thus ensures BPS saturation, but it would be interesting
to check by an explicit (though laborious) calculation that there
are no more composite operator counterterms in the renormalization
of the full susy algebra, and to find the superspace formulation from
which this follows.

To determine $Z_{\rm impr}$ we decompose $U=U_3=\int \6_i(S_3^a
\2\epsilon^{ijk}F_{jk}^a)d^3x$ as $U^{\rm impr}-\Delta U^{\rm impr}$
where
\bea
U^{\rm impr}&=&{2\03}\left[U-\int {i\016}\6_i (\epsilon^{ijk}
\bar \lambda \alpha^1 \gamma_{jk} \lambda) d^3x \right]\\
\Delta U^{\rm impr}&=&-{1\03}\left[U+\int {i\08}\6_i (\epsilon^{ijk}
\bar \lambda \alpha^1 \gamma_{jk} \lambda) d^3x \right]\equiv
-{1\03}(U+F).
\nonumber
\eea
For the one-loop composite operator renormalization of $U$
we thus have to consider four classes of proper diagrams:
graphs with the bosonic $U$ or the fermionic $F$ inserted
in proper one-loop diagrams with external bosonic background
fields or external fermionic fields.
The number of potentially divergent graphs with bosonic-bosonic
structure is 31, with fermionic-fermionic struture it is 2,
while there is one graph with $U$ insertion and external fermions,
and 3 graphs with $F$ insertion and external bosonic fields.
Of the set of 31 graphs, some vanish by themselves, some vanish
in the background covariant $R_{\xi=1}$ gauge, some subsets of
diagrams are finite, and if we only consider graphs with one external
field $S_3^3$ and one external gauge field $A_\mu^3$, only one
graph survives!

The set of one-loop graphs to be evaluated and their divergent
contributions (obtained in the topologically trivial sector) is
as follows (denoting gauge fields $A_\mu$ by wavy lines, the scalar field
$S_3$ by a dashed line, and fermions $\lambda$ by straight lines),

\vspace{5mm}
{\hspace*{6mm}$U\hspace{20mm}U\hspace{19mm}F\hspace{20.5mm}F\hspace{21mm}F$}
\centerline{
\includegraphics{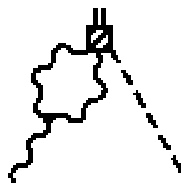}
\includegraphics{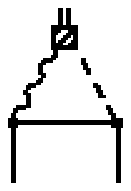}
\includegraphics{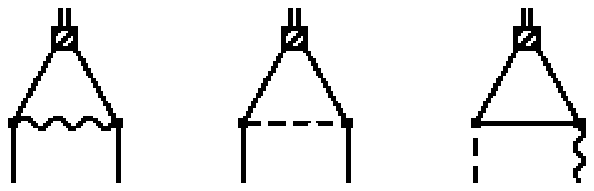}
}

\vspace{-10mm}
\bea
\nonumber&&\phantom{mm}\underbrace{\phantom{mmmmmmmmmmmm}}\\
\nonumber-4g^2UI \hspace{12mm} -4g^2FI &&\hspace{20mm} O(\xi-1) \hspace{20mm} -8g^2UI
\eea
to which one needs to add a term 
\be
 Z_\lambda F=-8g^2FI+O(\xi-1),
\label{graph}
\ee
since in $N=4$ theory there is nontrivial ordinary
wave function renormalization $Z_\lambda$ for fermions,
whereas bosonic fields do not obtain ordinary wave function renormalization 
in the gauge $\xi=1$.

This leads to the ($\xi$-independent) composite operator
counterterms,
\be
\Delta U_{c.o.r.}=4 g^2 (U+F)I,\quad
\Delta F_{c.o.r.}=8 g^2 (U+F)I.
\ee
From this it follows that $Z_{\rm impr}=1+12 g^2 I$, while
the improved part of $U$, $U^{\rm impr}=\frac23 U-\frac13 F$,
is indeed finite.

In the monopole background, the classical values of $-U$ and $M$
are equal and given by $4\pi m/g^2$ without renormalization of $m$ or
$g$. Since $U=U^{\rm impr}-\Delta U^{\rm impr}$, composite operator
renormalization produces the contribution
$12 g^2 I\times (+\frac{1}{3})\times 4\pi m/g^2=16 \pi m I$
to $-U$ as well as $M$, which indeed
cancels the divergence obtained above in (\ref{M1N4div}).
Thus we have $M^{(1)}=-U^{(1)} (\;=0)$ and BPS saturation is verified.

Having found that composite operator renormalization is needed for
the $N=4$ case, we should of course go back to the $N=2$ calculation
(and also all other one-loop calculations for solitons performed so far),
and make sure that in these cases there is no new contribution
that could upset the BPS saturation obtained previously.
For the kink in 1+1 dimensions, it is easy to check that
there are no divergent one-loop diagrams for composite operator
renormalization of the stress tensor (but it turns out that
there is in fact a need for composite operator renormalization
in the local energy density of 3+1 dimensional kink domain walls,
which does however not contribute to the integrated total energy
\cite{RSV}). For the vortex in 2+1 
dimensions\cite{Vassilevich:2003xk,Rebhan:2003bu,Olmez:2008au},
all currents are finite in dimensional regularizaton.
But for the $N=2$ monopole, no composite operator renormalization is needed for two reasons: (1) the improvement terms of the central charge $\Delta\mathcal{U}_{{\rm N=2},{\rm impr}}$ are proportional to the central charge $\mathcal{U}_{{\rm N=2}}$ itself and (2) the central charge $\mathcal{U}_{{\rm N=2}}$ is a finite operator due to ordinary renormalization. To prove these statements consider the central charge density for the $N=2$ model
\be
\mathcal U_{N=2}=
\2\epsilon_{ijk}\6_i(S^a F_{jk}^a)+\6_i(P^aF_{i0}^a)
=\mathcal U-\tilde{\mathcal U}.
\ee
The improvement terms in the susy current are given by
\bea
\Delta j_{N=2,\rm impr}^{\mu}&=&-\frac23 \Gamma^{\mu\nu}\6_\nu
[(P \Gamma^5+S\Gamma^6)\lambda]\nn
&=&-\frac23 \gamma^{\mu\nu}\6_\nu
[(P \gamma^5+iS)\lambda].
\eea
The susy variation of $\Delta j_{N=2,\rm impr}^{\mu}$ yields
$\Delta U_{N=2,\rm impr}$. Using $\delta P=\delta A_5
=\5\lambda \Gamma^5\epsilon=\5\lambda \gamma_5\epsilon$,
$\delta S=\delta A_6=\5\lambda \Gamma_6\epsilon=i\5\lambda\epsilon$
and $\delta\lambda=\2\Gamma^{PQ}F_{PQ}\epsilon$ we find
\bea
\Delta j_{N=2,\rm impr}^{0}&=&\gamma^{0j}\6_j
[\gamma_5\lambda(\5\lambda\gamma_5\epsilon)-\lambda(\5\lambda\epsilon)
+(P\gamma_5+iS)\2\gamma^{\rho\sigma}F_{\rho\sigma}\epsilon]\nn
&=&\gamma^{0j}\6_j
[-\4(\5\lambda \mathcal O^I\lambda)(\gamma_5 \mathcal O_I\gamma_5-
\mathcal O_I)\nn
&&\qquad\qquad +P\gamma_5\gamma^{0k}F_{0k}+iS\2\gamma^{kl}F_{kl}]\epsilon\nn
&\sim& \gamma_5\6_j(P F_{0j}-\2\epsilon^{jkl}SF_{kl})\epsilon,
\eea
because the terms with $\mathcal O_I\sim \gamma_{kl}$ cancel in the first
line. Thus $\Delta\mathcal{U}_{{\rm N=2},{\rm impr}}$ is proportional to $\mathcal{U}_{{\rm N=2}}$, so there are no fermionic terms in $\Delta\mathcal{U}_{{\rm N=2},{\rm impr}}$. Both $\mathcal U$ and $\tilde{\mathcal U}$ produce divergence proportional to $F$, but their sum cancels. Since ordinary renormalization already gave counterterms
which make $U_{N=2}$ finite, we do not need composite operator
counterterms, so we can set $Z_{\rm impr}=1$, leaving the
results for the $N=2$ monopole unchanged.

\section{Conclusions} 

The one-loop corrections to the mass and central charge of kinks, vortices (not discussed here, but treated in \cite{Vassilevich:2003xk,Rebhan:2003bu,Olmez:2008au}), and monopoles in $N=2$ and $N=4$ super Yang-Mills theory satisfy the BPS bound. To obtain this result, we needed to carefully regularize the susy field theories, which in our choice of regularization scheme meant that we needed to take extra dimensions into account. In these extra dimensions the modes of bosonic and fermionic quantum fields had extra momenta, and the square of these extra momenta gave an extra contribution to the 1-loop central charge. In addition we found that boundary terms contributed to the mass of the $N=4$ monopole. These boundary terms were divergent, and we needed multiplicative composite operator renormalization of the improvement terms in the stress tensor to obtain a finite quantum mass. The same composite operator renormalization was needed for the central charge. For the $N=2$ monopole, all boundary terms canceled, and there was no composite operator renormalization, but the sum over zero point energies in the bulk was divergent, and standard renormalization counter terms cancelled these divergences. For the susy kink, boundary terms could not even appear because the classical kink solution falls off exponentially fast.

We found that the 1-loop corrections to the susy kink and $N=2$ monopole are nonzero. In the literature it was assumed, or proofs were proposed, that these corrections vanish. Our results for the $N=2$ monopole agree with results based on holomorphicity by Seiberg-Witten\cite{Seiberg:1994rs}, which also require a nonvanishing correction to the mass and central charge (although this was noticed only subsequently\cite{Rebhan:2004vn}). This raises the question whether our results are consistent with Zumino's general proof that the sum over zero-point energies must vanish in any susy theory\cite{Zumino:1974bg}. This proof is based on path integrals and does not take into account regularization. Hence, it is not clear that there is a disagreement. There is a way of understanding our nonvanishing result. If one encloses the kink in a large box, and imposes susy boundary conditions, one finds a spurious boundary energy which one must subtract to obtain the true mass of the susy soliton.\cite{Shifman:1998zy}  Dimensional regularization by itself subtracts this spurious boundary energy. 

A superspace treatment of solitons would be useful, but we have found problems in gauge theories with a superspace $R_{\xi}$ gauge if solitons are present \cite{Goldhaber:2004kn}. A superspace treatment of the anomalies in the superconformal currents of the kink has been given in collaboration with Fujikawa\cite{Fujikawa:2003gi,Fujikawa:2003nm}, see also Shizuya\cite{Shizuya:2003vm,Shizuya:2004ii,Shizuya:2005th,Shizuya:2006et}. 

Our methods could perhaps be applied to D-branes\cite{Polchinski:1995mt}, at least the D-branes that are solitons. Also extension to finite temperature is interesting; in fact, we have found new surprises for kink domain walls at finite temperature\cite{RSV}. 

\section*{Acknowledgements}

We thank Yu-tin Huang for assistance in writing up this review
and acknowledge financial support from the Austrian Science Foundation FWF, project nos.\ J2660-N16 and
P19958.


\end{document}